\documentclass[10pt]{revtex4}
\usepackage{hyperref}
\usepackage{amsmath}
\usepackage[psamsfonts]{amssymb }
\usepackage{mathrsfs}
\usepackage{color}
\begin{document}

\title{Field-dependent quantum gauge transformation}

\author{ Sudhaker Upadhyay}
 \email {  sudhakerupadhyay@gmail.com; 
 sudhaker@boson.bose.res.in}

\affiliation { S. N. Bose National Centre for Basic Sciences,\\
Block JD, Sector III, Salt Lake, Kolkata -700098, India. }

\begin{abstract}
 In this paper we generalize the
 quantum gauge transformation of Maxwell theory obtained through gaugeon formalism.
 The generalization is made by making the bosonic transformation parameter
 field-dependent. The Jacobian of vacuum functional under 
 field-dependent quantum gauge transformation is calculated explicitly.
 We show that the quantum gauge transformation with a particular choice of  field-dependent  parameter connects the gaugeon actions of Maxwell theory in   two different gauges.
 We establish the result by connecting two well-known gauges, namely,   Lorentz gauge and axial gauge.
  \end{abstract}
\maketitle

\section{  Introduction}
Gauge theories are  the most successful theories  explaining the dynamics of elementary particles
and play   a  very  crucial  role in the unification of fundamental interactions.
 Being first in all the gauge theories,  the Maxwell theory of   electromagnetism  has become one of the main pillars of modern theoretical physics   playing the key role in the formulation and the development of Einstein's special theory of relativity. Quantum electrodynamics (QED) is an extension of Maxwell theory
describing the electromagnetic interaction of the nature.

In the quantization of  QED,  people generally consider the
gauge symmetry at the classical level   but not at the quantum level  because the quantum theory can be defined {\it properly} only after fixing the gauge. However, being gauge-fixed  the theory does not remain the (local) gauge invariant. The quantum gauge transformation for QED was first studied by Yokoyama utilizing a different 
  formalism which is commonly known as gaugeon formalism \cite{yo0}. Yokoyama  gaugeon formalism   provides a wider framework to quantize the general gauge theories \cite{yo0,yok,yo1,yo2,yo3}.
{The main idea  behind the gaugeon formalism  is to study the quantum gauge freedom}  by extending the configuration space with the introduction of some set of extra fields, so-called gaugeon fields, in the effective theory. Since,
the  gaugeon fields do not appear in the physical processes. Therefore, they are not the physical fields.
In fact, the gaugeon fields yield negative normed states in the theory which results the 
negative probability \cite{yo0}.
Hence, one needs to remove the unphysical modes present in the theory associated with 
gaugeon fields. Firstly,  the
  Gupta-Bleuler type subsidiary condition was implemented to remove the unphysical gaugeon modes. 
But this Gupta-Bleuler type subsidiary condition had been founded certain limitations \cite{yo0}. 
Further, the properties of BRST symmetry were utilized to get rid of such limitations of physical subsidiary conditions  by replacing them into a single Kugo-Ojima-type condition \cite{ki,mk,kugo,kugo1}. Along with BRST symmetry the extended gaugeon action  posses  the quantum gauge transformation under which the action remains \textit{form} invariant. 
Incidentally, Hayakawa and Yokoyama have founded that
{within} renormalization procedure the
  gauge parameters get  shifted from their original values \cite{haya}. However, within the framework of gaugeon formalism  a shift  in  
gauge parameter occurs naturally which gets identified with the renormalized parameter \cite{yo0}. 
The geugeon formalism  has already been utilized  by many people {in the context of various}   gauge theories  \cite{  mk1, naka, rko, miu, mir1, mir2,sud}.
Recently, the generalization of BRST symmetry  has analysed in the context 
of geugeon formalism \cite{sudhaker}.

Although the generalization of the BRST transformation by making the transformation
parameter finite and field-dependent has been studied extensively 
\cite{sdj,sud00, sb,sdj1,rb,susk,jog,sb1,smm,fs,rbs,sudha,rs}, but
the generalization of quantum gauge transformation in the similar fashion has not  yet been investigated.
Therefore, we take this opportunity to generalize the quantum gauge transformation
of Maxwell theory within gaugeon formalism.
Remarkable difference in these symmetry transformations
is that the transformation parameters of the BRST transformation and the quantum gauge transformation follow different statistics. For instance, the  parameter of the BRST transformation is fermionic in nature, however,  the parameter of the quantum gauge transformation is bosonic in nature. Therefore, the novelty of the present work is to
generalize the quantum gauge transformation by making the bosonic parameter
field-dependent. 

In the present work,   we first emphasize the effective Maxwell theory
   {analysing} the quantum gauge symmetry  through Yokoyama gaugeon formalism. For this purpose, we extend the configuration 
space by introducing the
gaugeon fields and corresponding ghost fields. 
Then, we investigate the quantum gauge symmetry for the extended Maxwell action, which we call the gaugeon-Maxwell action,  {incorporating} some extra 
quantum fields.
Furthermore, the quantum gauge symmetry is generalized by making the transformation parameter
field-dependent. The resulting field-dependent quantum gauge transformation leads to the 
the non-trivial field-dependent Jacobian within the functional integral. 
We compute the  Jacobian of the field-dependent 
quantum gauge transformation explicitly. 
Remarkably, for a particular choice of bosonic
field-dependent parameter the Jacobian changes the gaugeon-Maxwell action
from Lorentz gauge to axial gauge.
Although we establish the results with the help of an specific example, but these
hold for any arbitrary gauge.

This paper is organized in the following manner.
In section II, we  discuss the quantum gauge transformation for the Maxwell theory
within the framework of  Yokoyama gaugeon formalism.
Furthermore, in section III, we generalize the quantum gauge transformation by making the
transformation parameter field-dependent. 
The  novelty of such field-dependent quantum gauge transformation
is described in section IV.
We summarize  the present work in the last section. 
 
\section{Maxwell theory in Gaugeon formalism}
In this section, we discuss the
Yokoyama gaugeon formalism  {analysing} the quantum gauge freedom for Maxwell theory in covariant and non-covariant gauges.
In order to achieve the goal, we begin with  the effective action for 
Maxwell theory in Lorentz (covariant) gauge defined by  
\begin{eqnarray}
S^L_{M} =\int d^4x\left[-\frac{1}{4}F_{\mu\nu}F^{ \mu\nu  } -A_\mu\partial^\mu B 
+ \frac{\lambda}{2}  {B}^2
-i\partial^\mu c_\star \partial_\mu c\right], \label{ym}
\end{eqnarray}
where  $F_{\mu\nu}$  is the usual antisymmetric  field-strength tensor for the gauge field $A_\mu$. Here $B, c$ and $c_\star$  are the   
multiplier (auxiliary) field, the Faddeev-Popov ghost field and the anti-ghost field  respectively.
 
However, in accordance with above, the effective Maxwell action described in the axial (non-covariant) gauge is defined by
\begin{eqnarray}
S^A_{M} =\int d^4x\left[-\frac{1}{4}F_{\mu\nu}F^{ \mu\nu} -A_\mu\eta^\mu B
+ \frac{\lambda}{2}   {B}^2
-i\eta^\mu c_\star \partial_\mu c\right], \label{ym1}
\end{eqnarray}
where $\eta_\mu$ is an  arbitrary constant four vector.
The effective actions $S^L_{M}$ and $S^A_{M}$ are invariant under the following
nilpotent  BRST transformations (i.e. $\delta_b^2=0$):
 \begin{eqnarray}
 \delta_b A_\mu &=& -\partial_\mu c\ \eta,\  \
 \delta_b c=0,\nonumber\\
 \delta_b c_\star &=&iB \  \eta,\ \ \ \ \
 \delta_b B=0,
 \end{eqnarray}
where $\eta$ is an infinitesimal, anticommuting but global parameter.

Now, the   gaugeon effective action  {corresponding to} the Maxwell theory (\ref{ym}) is 
obtained by introducing 
  the gaugeon field $Y$ and its associated 
field $Y_\star$ (both subjected
to the Bose-Einstein statistics)  as follows \cite{kosaki}
\begin{eqnarray}
S^L_{Y} =\int d^4x\left[-\frac{1}{4}F_{\mu\nu}F^{ \mu\nu} -A_\mu\partial^\mu B
+\partial_\mu Y_\star \partial^\mu Y +\frac{\varepsilon}{2} (Y_\star +\alpha B)^2
-i\partial^\mu c_\star \partial_\mu c\right],\label{ga}
\end{eqnarray}
where $\alpha$ denotes the group vector valued gauge-fixing parameter and $\varepsilon$ refers the ($\pm$) sign  factor.
The
quantum gauge transformation,  which leaves the quantum action  (\ref{ga})  form invariant, is demonstrated as
\begin{eqnarray}
&&A_\mu\longrightarrow \hat A_\mu = A_\mu +  \alpha\partial_\mu Y\tau,\nonumber\\
 &&B\longrightarrow \hat B = B,\nonumber\\
 &&Y_\star\longrightarrow \hat Y_\star = Y_\star - \alpha B \tau,\nonumber\\
 &&Y\longrightarrow \hat Y =Y,\nonumber\\
  &&c\longrightarrow \hat c =c,\nonumber\\
  &&c_\star\longrightarrow \hat c_\star =c_\star, \label{ko}
\end{eqnarray}
where     $\tau$
is a bosonic transformation parameter.
The form invariance of quantum action  (\ref{ga}), under the above quantum gauge transformation, reflects a natural shift  in parameter $\alpha$
as following
\begin{equation}
\alpha\longrightarrow \hat\alpha =\alpha+ \alpha\tau.\label{alp}
\end{equation}
Furthermore, we note that the gaugeon fields are not the physical fields and, therefore, to define physical states we need  to remove the unphysical modes associated with them.
This can be achieved by  imposing the following
Gupta-Bleuler type subsidiary condition:
 \begin{eqnarray}
   (Y_\star +\alpha B)^{(+)}|\mbox{phys}\rangle &=&0.\label{con}
 \end{eqnarray}
 This   Gupta-Bleuler
  condition removes the unphysical  gaugeon mode. 
However, the unphysical modes associated with the gauge field are removed by 
utilizing the Kugo-Ojima type restriction. The Kugo-Ojima type restriction is valid for all kind of theories but, having certain limitations, the Gupta-Bleuler
type  subsidiary condition is not. For example,
 the decomposition of combination $(Y_\star +\alpha B)$ in positive and negative frequency parts can be done only if   the combination  satisfies the following free equation:
\begin{equation}
\square (Y_\star +\alpha B) =0,
\end{equation} 
where $\square=\partial_\mu\partial^\mu$.
Furthermore,  such kind of limitation is   improved by introducing the Faddeev-Popov ghosts $K$ and $K_\star$ corresponding to  the
gaugeon fields $Y$ and $Y_\star$ in the
 Yokoyama effective action (\ref{ko})  as follows:
\begin{eqnarray}
S^L_{YB} =\int d^4x\left[-\frac{1}{4}F_{\mu\nu}F^{ \mu\nu} -A_\mu\partial^\mu B
+\partial_\mu Y_\star \partial^\mu Y +\frac{\varepsilon}{2} (Y_\star +\alpha B)^2
-i\partial^\mu c_\star \partial_\mu c -i\partial^\mu K_\star \partial_\mu K\right].\label{gan} 
\end{eqnarray}
Now, the effective action, $S^L_{YB}$, admits the following nilpotent BRST symmetry transformations:
 \begin{eqnarray}
 \delta_b A_\mu &=& -\partial_\mu c\ \eta,\  \
 \delta_b c=0,\nonumber\\
 \delta_b c_\star &=&iB  \  \eta,\ \ \ \ \
 \delta_b B=0,\nonumber\\
  \delta_b Y &=&-K \  \eta,\ \ \
 \delta_b K=0,\nonumber\\
  \delta_b K_\star &=& -iY_\star \  \eta,\ \
 \delta_b Y_\star =0.\label{brst}
 \end{eqnarray}
Using the Noether's theorem we calculate the  conserved charge corresponding to the    BRST symmetry  (\ref{brst}) as  follows
\begin{eqnarray}
Q =\int d^3x \left[-F^{ 0\nu  } \partial_\nu c 
- \dot c  B  -Y_\star  \dot K  \right],
\end{eqnarray}
which  helps  in replacing the two Yokoyama subsidiary conditions (namely, Kugo-Ojima type and  Gupta-Bleuler type (\ref{con}))  by a single
  Kugo-Ojima type condition (for details see, e.g., \cite{mk}). 
 The effective action (\ref{gan})  also admits the following 
 quantum gauge transformations ($\delta_q$):
 \begin{eqnarray}
&&\delta_q A_\mu  =   \alpha\partial_\mu Y\tau,\ \ \
 \delta_q B =  0,\nonumber\\
 &&\delta_q Y_\star  =  -\alpha B \tau,\ \ \
  \delta_q Y  =0,\nonumber\\
  &&\delta_q c = K \tau,\ \
 \delta_q c_\star = 0, \ \
  \delta_q K=0,\nonumber\\
  &&\delta_q K_\star =-c_\star \tau,\ \ \delta_q  \alpha =\alpha\tau, \label{fil}
\end{eqnarray}
where $\tau$ is bosonic transformation parameter.
Here we observe that these transformations are also nilpotent in nature, i.e. $\delta_q^2=0$.

Now, we define the path integral for the gaugeon-Maxwell theory in Lorentz gauge
described by the action (\ref{gan})
as follows:
\begin{eqnarray}
Z^L[0] =\int {\cal D}\phi\ e^{iS^L_{YB}[\phi]},\label{zen}
\end{eqnarray}
where the generic field $\phi$  refers all the  fields collectively.
Following the similar steps discussed above
for the Lorentz gauge case, the path integral for the gaugeon-Maxwell theory
in axial gauge is defined by
\begin{eqnarray}
Z^A[0] =\int {\cal D}\phi\ e^{iS^A_{YB}[\phi]},\label{zen1}
\end{eqnarray}
with the effective gaugeon-Maxwell action in axial gauge,
\begin{eqnarray}
S^A_{YB} &=&\int d^4x\left[-\frac{1}{4}F_{\mu\nu}F^{ \mu\nu} -A_\mu\eta^\mu B
+\eta_\mu Y_\star \partial^\mu Y +\frac{\varepsilon}{2} (Y_\star +\alpha B)^2\right.\nonumber\\
&-&\left. i\eta^\mu c_\star \partial_\mu c -i\eta^\mu K_\star \partial_\mu K\right].\label{gan1}  
\end{eqnarray}
The effective action $S^A_{YB}$ is also  invariant under the BRST transformations (\ref{brst}).
\section{Field-dependent quantum gauge transformation}
In this section, we investigate the methodology of the field-dependent 
quantum gauge transformation characterized by the field-dependent bosonic  parameter.
To achieve the goal, we first define the general nilpotent
quantum gauge transformation for the generic field $\phi_\alpha(x)$, written compactly, as
 \begin{eqnarray}
\delta_q \phi_\alpha(x)=\phi_\alpha'(x)-\phi_\alpha(x)={\cal R}_\alpha(\phi) \tau,
 \end{eqnarray}
where ${\cal R}_\alpha(\phi)$ is the generic variation of field $\phi_\alpha(x)$ 
under the quantum gauge transformation satisfying $\delta_q{\cal R}_\alpha(x)=0$
 and $\tau$ is the parameter of
transformation satisfying Bose-Einstein statistics.

Now, we propose the  field-dependent quantum gauge transformation defined by
 \begin{eqnarray}
 \delta_q\phi_\alpha(x)=\phi_\alpha'(x)-\phi_\alpha(x)={\cal R}_\alpha(\phi) \tau[\phi],\label{qg}
 \end{eqnarray}
 where parameter of transformation $\tau[\phi]$ depends on fields explicitly.
 Now, it is obvious that such  field-dependent quantum gauge transformations
do not remain nilpotent any more.
In spite of being non-nilpotent  the field-dependent quantum gauge transformation (\ref{qg})
leaves the quantum action $S^L_{YB}$ given in (\ref{gan}) form invariant. However,
the functional measure defined in (\ref{zen}) is not invariant under such field-dependent quantum gauge transformation.
If we apply the  field-dependent  quantum gauge transformation defined in (\ref{qg}) on the generating functional (\ref{zen}), the generating functional gets transformed   as follows
\begin{eqnarray}
\delta_q Z^L[0] &=&\int {\cal D}\phi ( \mbox{Det} J[\phi]) e^{iS^L_{YB}[\phi]},\nonumber\\
 &=&\int {\cal D}\phi\  e^{i\left(S^L_{YB}[\phi]-i \mbox{Tr} \ln J[\phi]\right)}.
 \label{zl} \end{eqnarray}
 Furthermore, we calculate the Jacobian matrix of field-dependent
 quantum gauge transformation (\ref{qg}) as 
 \begin{eqnarray}
J_\alpha^{\ \beta}[\phi]= \frac{\delta \phi'_\alpha}{\delta\phi_\beta}&= &\delta_\alpha^{\ \beta}+
\frac{ \delta{\cal R}_\alpha(\phi)}{\delta\phi_\beta} \tau[\phi] +
{\cal R}_\alpha(x)\frac{ \delta\tau[\phi]}{\delta\phi_\beta},\nonumber\\
&= &\delta_\alpha^{\ \beta}+
{\cal R}_\alpha^{\ ,\beta}(\phi)  \tau[\phi] +
{\cal R}_\alpha(\phi)\tau^{,\beta}[\phi].\label{jac}
 \end{eqnarray}
Utilizing the nilpotency property of quantum gauge transformation (i.e. $\delta_q{\cal R}_\alpha(\phi)=0$) and relation (\ref{jac}),
we compute
 \begin{eqnarray}
 \mbox{Tr} \ln J[\phi] &=& {\sum_{n=1}^{\infty} \frac{(-1)^{n+1}}{n}\mbox{Tr} (
{\cal R}_\alpha^{\ ,\beta}   \tau +
{\cal R}_\alpha \tau^{,\beta} )^n },\nonumber\\
&=& { \sum_{n=1}^{\infty} \frac{(-1)^{n+1}}{n}  (
\delta_q \tau^{ \alpha}[\phi]  )^n} ,\nonumber\\
 &=&\ln(1+\delta_q \tau[\phi]),\label{J}
 \end{eqnarray}
 where  $\tau[\phi]$ is considered up to linear order which reflects the infinitesimal 
nature of parameter even though it depends on the fields explicitly. 
 Consequently, the expression (\ref{zl}) reads
 \begin{eqnarray}
\delta_q Z^L[0] =\int {\cal D}\phi\  e^{i\{ S^L_{YB}[\phi]-i \ln(1+\delta_q \tau[\phi]) \}},\label{dzl}
 \end{eqnarray}
which is nothing but the expression of generating functional for
gaugeon-Maxwell theory having an additional term, {$-i\ln (1+\delta_q \tau[\phi])$, in the effective action due to the Jacobian}.
 Therefore, we conclude that under the field-dependent quantum gauge transformation the
 generating functional changes from one effective action to another.
\section{Connecting different gauges of gaugeon-Maxwell action}
In this section,
we explicitly mention the remarkable features of the field-dependent quantum gauge transformation  
which  connects the two different gauges of gaugeon-Maxwell theory  for a particular choice of 
the field-dependent parameter.
In this regard, we start by making the
quantum gauge transformation defined in (\ref{fil}) field-dependent  as follows:
 \begin{eqnarray}
&&\delta_q A_\mu  =   \alpha\partial_\mu Y\tau[\phi],\ \ \
 \delta_q B =  0,\nonumber\\
 &&\delta_q Y_\star  =  -\alpha B \tau[\phi],\ \ \
  \delta_q Y  =0,\nonumber\\
  &&\delta_q c = K \tau[\phi],\ \
 \delta_q c_\star = 0, \ \
  \delta_q K=0,\nonumber\\
  &&\delta_q K_\star =-c_\star \tau[\phi],\ \ \delta_q  \alpha =\alpha\tau[\phi],
\end{eqnarray}
where $\tau[\phi]$ is {the} field-dependent transformation parameter. Here the
specific choice  of the field-dependent transformation parameter
is made by
\begin{eqnarray}
\tau[\phi] &=&-{\int d^4 x}\ \alpha Y_\star B  (\alpha
B)^{-2} \left[\exp\{i(A_\mu\partial^\mu B
-\partial_\mu Y_\star \partial^\mu Y
+i\partial^\mu c_\star \partial_\mu c  +i\partial^\mu K_\star \partial_\mu K 
\right.\nonumber\\
&-&\left. A_\mu\eta^\mu B
+\eta_\mu Y_\star \partial^\mu Y
-i\eta^\mu c_\star \partial_\mu c -i\eta^\mu K_\star \partial_\mu K)\}  -1\right].
\end{eqnarray}
Now, the expression (\ref{J}) for the above  field-dependent transformation parameter
yields
\begin{eqnarray}
\ln(1+\delta_q\tau[\phi]) &=&i\int d^4 x\ (A_\mu\partial^\mu B
-\partial_\mu Y_\star \partial^\mu Y
+i\partial^\mu c_\star \partial_\mu c  +i\partial^\mu K_\star \partial_\mu K 
\nonumber\\
&-& A_\mu\eta^\mu B
+\eta_\mu Y_\star \partial^\mu Y
-i\eta^\mu c_\star \partial_\mu c -i\eta^\mu K_\star \partial_\mu K).
\end{eqnarray}
With this identification, the expression  (\ref{dzl}) gets the
following form:
\begin{eqnarray}
\delta_q Z^L  &=& \int {\cal D}\phi\ \exp\left[i\{ S^L_{YB}[\phi]+\int d^4 x\ (A_\mu\partial^\mu B
-\partial_\mu Y_\star \partial^\mu Y
+i\partial^\mu c_\star \partial_\mu c  +i\partial^\mu K_\star \partial_\mu K 
\right.\nonumber\\
&-&\left. A_\mu\eta^\mu B
+\eta_\mu Y_\star \partial^\mu Y
-i\eta^\mu c_\star \partial_\mu c -i\eta^\mu K_\star \partial_\mu K )\}\right].\nonumber\\
&=&\int {\cal D}\phi\ e^{iS^A_{YB}[\phi]}=Z^A,
 \end{eqnarray}
 which is nothing but the generating functional for 
 gaugeon-Maxwell theory in axial gauge.   Hence, the field-dependent quantum gauge transformation  enables one to go back and forth between
the two sets of gauges.
 We, therefore, end up  the section with the remark that within gaugeon formalism  the quantum gauge transformation with a particular choice of the
 field-dependent parameter maps 
 the path integrals of the same effective theory in two different gauges.

\section{  Conclusions}
In this paper, we have discussed the Maxwell theory in gaugeon formalism to analyse the
quantum gauge freedom in great details. 
In the framework of gaugeon formalism, we have 
investigated the quantum gauge transformation characterized 
by an infinitesimal bosonic parameter which leaves the quantum action \textit{form} invariant. 
Under the quantum gauge transformation a natural shift in gauge parameter
has been observed. Furthermore, we have constructed  
{the} gaugeon-Maxwell action in two different gauges (namely, in the Lorentz
  and   the axial 
gauges) possessing the BRST as well as the quantum gauge symmetries.
The infinitesimal bosonic parameter of {the} quantum
gauge transformation  has been made field-dependent.
Furthermore, the Jacobian of path integral measure under {the} field-dependent
quantum gauge transformation has been computed.
Remarkably, we have observed that under {the} field-dependent transformation
with specific bosonic field-dependent parameter the generating functional
of gaugeon-Maxwell theory changes from the Lorentz gauge to the axial gauge.
  Although there are many choices of the gauge condition but the physical quantities {do} 
  not depend on any of them. Therefore, the spectrum of the physical theory remains unaltered under 
such field-dependent quantum gauge transformation.

We have made all the computations with the source free partition functions.
However, it would be possible to make such analysis for partition functions having an external source. In my opinion, for {the} partition functions having an
 external source such analysis
will connect the propagators corresponding to the appropriate gauges because the connection
of propagators in {the} Lorentz and {the} axial gauges under finite field-dependent BRST transformation had already been established \cite{jog1}.
Also, there are  not any ambiguities in dealing with the singularities of the propagators corresponding 
to the Lorentz and the axial gauges. Naturally, a large number of the
practical as well as the formal calculations have been made in Lorentz gauges. 
The main disadvantage of Lorentz gauge choice in non-Abelian gauge theory is, however, that it requires a ghost action which complicates the calculations. For this reason,
another set of gauge (namely  the axial gauge)  has often been founded favouring calculations. 
Therefore, it would be interesting to generalize the results in non-Abelian gauge theory, in
higher form gauge theories and in perturbative gravity theory as well.

\end{document}